# A Digital Twin for Reconfigurable Intelligent Surface Assisted Wireless Communication

Baoling Sheen, Jin Yang, Xianglong Feng, and Md Moin Uddin Chowdhury

***Abstract*—**Reconfigurable Intelligent Surface (RIS) has emerged as one of the key technologies for 6G in recent years, which comprise a large number of low-cost passive elements that can smartly interact with the impinging electromagnetic waves for performance enhancement. However, optimally configuring massive number of RIS elements remains a challenge. In this paper, we present a novel digital-twin framework for RIS-assisted wireless networks which we name it Environment-Twin (Env-Twin). The goal of the Env-Twin framework is to enable automation of optimal control at various granularities. In this paper, we present one example of the Env-Twin models to learn the mapping function between the RIS configuration with measured attributes for the receiver location, and the corresponding achievable rate in an RIS-assisted wireless network without involving explicit channel estimation or beam training overhead. Once learned, our Env-Twin model can be used to predict optimal RIS configuration for any new receiver locations in the same wireless network. We leveraged deep learning (DL) techniques to build our model and studied its performance and robustness. Simulation results demonstrate that the proposed Env-Twin model can recommend near-optimal RIS configurations for test receiver locations which achieved close to an upper bound performance that assumes perfect channel knowledge. Our Env-Twin model was trained using less than 2% of the total receiver locations. This promising result represents great potential of the proposed Env-Twin framework for developing a practical RIS solution where the panel can automatically configure itself without requesting channel state information (CSI) from the wireless network infrastructure.

## I. Introduction

RIS (Reconfigurable Intelligent Surfaces) has been envisioned as a promising technology to reduce the energy consumption and improve the communication performance by artificially reconfiguring the propagation environment of electromagnetic waves. As such, RISs have the huge potential to revolutionize the design of wireless networks, particularly when combined and integrated together with other 6G candidate technologies such as terahertz communications and AI-empowered wireless networks.

To fully realize the potential and benefits of RIS, there are three key requirements to allow scalable and low-cost deployments:

1) **Low overhead and real-time** control of RIS elements. This is important as we move towards more ultra-reliable low-latency communication (URLLC) driven applications, in conjunction with other types of traffic.
2) **Predictive** control of RIS elements. For high frequency band communication, line-of-sight (LOS) channel presence is an important assurance for high reliability communication needs. RIS can help "create" such LOS channel but only if we can predict the channel dynamics and configure RIS elements proactively.
3) **Minimum "burden"** or requirements on communication devices and nodes. We want the RIS panel to be as "simple and dumb" as possible (e.g. completely passive), allowing "standalone" deployment [1] without deep integration needed with existing communication systems. We would also like to see it works with devices which have limited or no signal processing capability, for example in the wireless power transfer use case of RIS.

It is apparent that RIS-assisted networks/systems, such as communication, sensing, wireless charging, etc. are complex and very challenging to design. This originates from the large number of parameters to be optimized based on the contextual information, and real-time decisions to be made every time that the network conditions change, e.g., channel condition changes, positions of the users change, etc. Traditional or analytics-based solutions build upon communication theories, mathematical models, and optimization algorithms. While solid and very successful, they face challenges in supporting RIS-assisted systems and meeting the increasingly stringent requirements for future applications.

To realize the optimal control of the RIS panels, one of the first and fundamental steps, in traditional approach, is to "sense" and acquire channel state information (CSI). As pointed

B. Sheen and J. Yang are with Futurewei Technologies, Bridgewater, NJ 08807 USA (e-mail: baoling.sheen@futurewei.com; jin.yang@futurewei.com).

X. Feng is with the Electrical and Computer Engineering Department, Rutgers University, New Brunswick, NJ 08901 USA (e-mail: xf56@scarletmail.rutgers.edu).

M. M. U. Chowdhury is with the Electrical and Computer Engineering Department, North Carolina State University, Raleigh, NC 27695 USA (e-mail: mchowdh@ncsu.edu).

out in [2] the CSI acquisition problem in a RIS-assisted system has its unique challenges. High-dimensional channel space introduced by the large number of RIS elements implies higher pilot training overhead (both the pilot bits required and the associated processing power) and longer delay. The passiveness of RIS poses extra challenges for signal processing algorithms. On top of that, the channel model of a RIS-assisted MIMO system has not yet been well understood. More critically, such "sensed" channel status is mostly "after fact" and the system can only react to what already happened, which makes it extremely hard to combat any unfavorable or sudden change of propagation conditions, especially at high frequency bands. Lastly, from the deployment perspective, such an approach would require the communication nodes and devices to support the necessary signal processing functions and low layer protocols of the communication systems. Such requirements impose constraints on candidate scenarios of RIS systems.

In this paper, we call for a new paradigm – ***environmental-twin (Env-Twin) Assisted Communication***, driven by two critical observations:

- Firstly, we are entering into a completely new space of "extreme" performance requirements for future communication applications as highlighted in [3], in terms of delay, reliability, throughputs, etc., while deploying higher frequency spectrum. Meanwhile, we are going to see a growing mixture of solutions including various cell layouts (macro cell, small cells, Terrestrial as well as Non-Terrestrial Networks), spectrum ranges (including millimeter wave (mmWave), terahertz, or even visible light communication), and a whole new dimension of channel dynamics enabled by RIS. The orchestration and optimal choices and configurations of these tools depend more and more on their "**close fitness**" to the specific underlying physical environment, e.g. surroundings and user distributions/mobility etc. Traditional communication technologies, however, assume few or zero knowledge of such environment. While being a great strength leading us this far, it faces growing challenges and limitations to find such "close fitness" to the underlying environment and meet higher performance demands.
- At the same time, we are at the beginning of a new era of industrial transformation enabled by sensing, digitization, and connectivity. Combined with the growing power of artificial intelligence / machine learning (AI/ML) algorithms, digital twins can be constructed for real-world entities, whether it be an organization, building, road, city, product, person, or process. Such digital twins allow us to observe the surrounding environments, the interactions, and impacts on the performance of the communication systems, discover rich knowledge and patterns, and finally make decisions at various granularity levels accordingly.

This new "Env-Twin aided communication" paradigm has great potential in advancing the communication technologies in many aspects. For the first time, the communication system can self-learn the relationships among high dimensional factors, which is hard to capture using traditional analytical approach, and leverage the discovered knowledge of the underlying environment and network behaviors for performance and other optimizations. In this paper we use an "RIS-embedded environment" as an example to study the feasibility and benefits of our proposed Env-Twin framework.

There are two groups of research activities that are closely related to our work. The first one is under the theme of "environmental aware communication". In [4], a "Channel Knowledge Map" is built to enable environmental aware communication especially in addressing the practical challenges brought by the drastically increased channel dimensions and training overhead. In [5], an "environmental AI" approach is envisioned in which (RIS control) agents interact with the environment through their embedded sensors, and learn the optimal policy via reinforcement learning, which needs to be optimally converged in a time period much shorter than the coherence time of the wireless environment. While we share the general view and motivations of the above approaches, in this paper we present a more systematic framework of a digital twin-based new paradigm to capture different aspects of an RIS-embedded environment and behavior leveraging state-of-art ML algorithms, without the need of fast-convergence as highlighted in [5].

Another group of related work is applying ML for RIS control [5]–[10] . These papers tend to use a machine learned model for a specific problem or scenario. We try to go beyond and propose a more general solution consisting of a set of models capturing various aspects of the environment and show how the learning efficiency can be dramatically improved by careful design of data representation and model architecture.

Within the context described above, the main contributions of this paper can be summarized as follows:

- We propose a systematic framework to construct an "Env-Twin" for communication systems, which is well beyond digital representations and simulation, including key functional components and machine learned models, leveraging either synthetic data or real-world observations.
- A case study and a simplified instance of such Env-Twin for RIS-embedded environment is carried out and we demonstrate how it can be used to achieve near optimal control of RIS configuration without explicit channel knowledge acquisition. We also demonstrate that by careful design of data representation and combination of both domain knowledge and AI techniques, we can achieve a very data-efficient learning compared to benchmark method.

The rest of this paper is organized as follows. In Section II, we first introduce the general framework of Env-Twin framework, then we describe the network model and the channel model for an RIS-embedded environment. Section III summarizes the problem space and the mathematic formulation. Note that although our proposed approach is ML-based, we describe the channel model and problem formulation based on traditional communication theory-based mathematic formulas in Section II and Section III; later in this paper, we describe how the relationships can be learned without using channel state information. In Section IV, we present our deep learning (DL)- based approach to learn an Env-Twin model, its architecture, loss function, and algorithm pseudo-code. Section V shows numerical analysis results for the proposed approach and comparison with the baseline approach. Finally, we present conclusion and share our view for future research topics in Section VI.

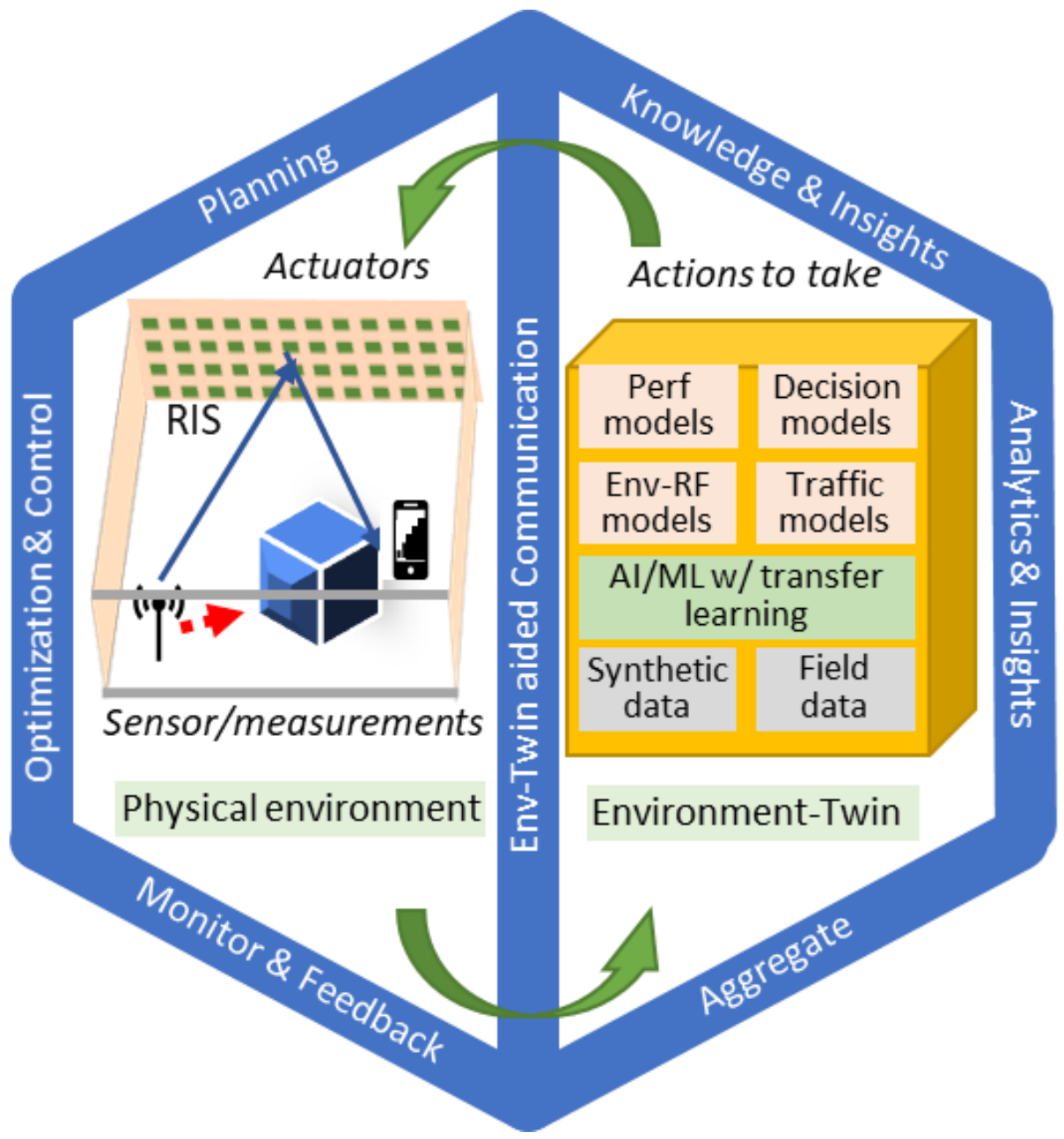

Fig. 1. A general framework of communication Env-Twin

**Notation:** Unless otherwise stated, we use the following notation throughout this article: scalars, vectors, and matrices are denoted by lower/upper case, boldface lower case, and boldface uppercase letters, respectively. For an arbitrary matrix $\mathbf{A}$, $\mathbf{A}^*$ and $\mathbf{A}^T$ are its conjugate and transpose, respectively. diag ($\mathbf{a}$) is a diagonal matrix created by putting the elements of $\mathbf{a}$ in its diagonal positions. vec($\mathbf{A}$) is a vector whose elements are the stacked columns of matrix $\mathbf{A}$. Hadamard product of matrices $\mathbf{A}$ and $\mathbf{B}$ is represented by $\mathbf{A}\odot\mathbf{B}$. $\mathcal{N}(\boldsymbol{\mu},\mathbf{R})$ is a complex Gaussian random vector with mean $\boldsymbol{\mu}$ and covariance matrix $\mathbf{R}$. Imaginary unit of a complex number is denoted by $j=\sqrt{-1}$. Finally, $\mathbb{E}[\cdot]$ denotes expectation.

## II. General Framework and System Model

### A. *A general framework for communication Env-Twin*

An overall Env-Twin framework in our design, as depicted in Fig. 1, consists of a set of models that represent layered abstractions of the underlying environment and behaviors, each of which captures the interaction and (ideally the) causal relationship, typically between ***States*** and ***Observations***, or between ***States*** and ***Decisions.***

**State** represents different aspects of the underlying communication environment. Firstly, the physical surroundings that impact RF propagation, which can be represented in 3D maps, and can be enriched with additional information like building surface materials, time-of-day, weather, etc. For the RIS-embedded environment, it includes each RIS panel position, size, and its element configurations. Secondly, the transmitter location, operating frequency band, power, and antenna pattern, etc. Finally, the receiver, e.g. user equipment (UE) or device, location, movement speed, pose, etc.

**Observations** are the measurements we can obtain to reflect the interactions of underlying impacting factors. There are two levels of observations. One is driven by immediate physical laws especially the observed RF channel characteristics e.g. the signal strength observed for a channel, or because of human/device movements, e.g. the traffic loading and mobility patterns/trajectories. The second level of observation is performance related (various levels of Key Performance Indicators (KPI's) that the industry is using, including the achievable bit rate) which is a result of many underlying factors including those reflected by the first level of observations.

Support **Decision** making at various granularity levels is the real goal of such Env-Twin. Decision engines (or models) can be learned from state-action-reward observations (e.g. via reinforcement learning) or derived from the state-observation relationship models if the actions can be observed or calculated as part of the state.

The decisions to be made can be at various granularity levels (spatial and time horizon). Typically, the finer the granularity (e.g. at the TTI interval), the more detailed and/or accurate information would be needed for state and observation measurements. At the same time, it is critical to apply domain knowledge to assess the prediction feasibility/power from state to observations or decisions.

There is no fixed set of models for an Env-Twin. The design and number of models depend on the applications and target problems, although we envision there will be a common set of models for typical application scenarios.

In general, an Env-Twin is valid for a particular surrounding setting which implies a different Env-Twin is to be constructed for a new environment. However, with a more general representation of environmental information and the advancements in ML techniques, like transfer learning, it is possible to derive an initial Env-Twin without training from scratch.

The key strength of such Env-Twin-based solution, compared with the traditional analytical approach, lies in the following aspects:

1) *Pattern and predictability*: Although highly dynamic in nature, wireless network does present repeatable patterns based on the surroundings, traffic, UE locations etc. In traditional approach, there is no mechanism and capability to capture such patterns, and its operation assumes no such knowledge available. With Env-Twin, the learned and abstracted patterns, in the forms of various models, can be used for prediction and decision making.
2) *Learning complex relationship with or without a model*: In traditional optimization method, an explicit relationship (model) is required among the measurements and target optimization function. For many problems, especially complicated system-level problems, such models are hard to craft. Env-Twin models, instead, can be learned with or without such explicit domain knowledge, or in a hybrid manner. Furthermore, impacts from "hidden" variables can be (partially) captured by other proxy features such as time-of-day, location, traffic loading etc.

3) *Multi-modal environmental observations*: In traditional method, channel status needs to be estimated explicitly, e.g. using pilots for FDD downlink channels. Env-Twin, on the other hand, can leverage any mode of sensing capabilities, both RF and non-RF (e.g. images), to associate a multi-modal state with corresponding channel property or other attributes. As illustrated in [11], such approach enhances the system predictability and performance.

For a RIS-embedded environment, its Env-Twin will capture the impacting relationship of RIS panels and configurations to the devices based on their locations and other attributes. Such relationship can be used for optimal control of RIS elements, placements of multiple panels and other RIS-aided applications, without explicit channel measurement/estimations, or combined with the traditional approach to improve the overall overhead and performance.

### B. An RIS-embedded Environment

In this subsection, we consider an RIS-embedded environment as the target physical world environment of the Env-Twin framework introduced in the previous subsection and as a use case we study in this paper. We describe the models (network model and channel model), and assumptions of such an environment first, followed by problem formulation in Section III. These models and formulas are used to construct/simulate the RIS-environment upon which observations are drawn to train the Env-Twin model.

- *Network Model*: We consider an RIS-assisted downlink communication system, where an RIS composed of $N$ passive reflecting elements is deployed to assist in the communication from the transmitter (Tx) to a receiver (Rx). Tx and Rx are both equipped with single omni-directional antennas each. Our model can be easily extended to a multi-antenna scenario.

  As done in [1][12], we adopt an OFDM-based system of $K$ subcarriers, and let $\boldsymbol{h}_{T,k},\ \boldsymbol{h}_{R,k} \in \mathbb{C}^{N\times 1}$ be the $N \times 1$ channels from Tx/Rx to RIS at the $k^{\text{th}}$ sub-carrier**.** In this paper, we will focus on the case where the direct link between the Tx and Rx does not exist i.e., the direct link is either blocked or has negligible received power compared to that received beside the RIS-assisted link and there is no other environmental input to the receiver besides the RIS. The interactions of the RIS elements on the incident signal is modeled by the diagonal matrix

$$\boldsymbol{\Theta}_k = \text{diag}\left(\alpha_1 \exp(j\theta_1), \dots, \alpha_\text{N} \exp(j\theta_N)\right), \quad (1)$$

  where $\theta_n \in [0,2\pi)$ and $\alpha_n \in [0,1]$ represent the phase-shift and the amplitude coefficient for element $n \in \{1,2,\dots,N\}$, respectively. According to [13], the amplitude coefficients are dependent on the reflection phase shifts. However, for simplicity, we assume no loss/attenuation on the incident signals i.e., $\alpha_n = 1$ for all elements in the sequel of the paper. The same phase shift will be used for all sub-carriers since phase shifting is done in the analog domain [12][14]. Hence, we will drop the subscript $k$ from $\boldsymbol{\Theta}_k$ in the rest of this paper.

  Then the received signal at the receiver can be expressed as:

$$y_k = \boldsymbol{h}_{R,k}^T \boldsymbol{\Theta} \boldsymbol{h}_{T,k} x_k + n_k, \quad (2)$$

$$\overset{(a)}{=} \left(\boldsymbol{h}_{R,k} \odot \boldsymbol{h}_{T,k}\right)^T \boldsymbol{v} x_k + n_k, \quad (3)$$

  where $\mathbf{v}$ is the RIS reflection beamforming vector i.e., $\boldsymbol{\Theta} = \text{diag}(\boldsymbol{v})$, $x_k$ denotes the transmitted signal over the $k^{\text{th}}$ subcarrier and satisfies $\mathbb{E}[|x_k|^2] = \frac{P_T}{K}$, with $P_T$ representing the total transmit power and $n_k \sim \mathcal{N}(0, \sigma_n^2)$ is the noise power. The phase of each RIS element can be adjusted through the PIN diodes [14], which are controlled by the RIS-controller over the backhaul link. Here we assume that the backhaul link uses separate frequency resource other than the data communication frequency. We also assume that the signals reflected by RIS two or more times are ignored due to the severe "distance-product" power loss over multiple reflections [14].

- *Channel Model:* Motivated by [1][12], here we also adopt the wideband geometric channel model [12] to model the channels $\boldsymbol{h}_{T,k},\ \boldsymbol{h}_{R,k}$ between the Tx/Rx and the RIS. Let us consider a Tx-to-RIS channel, $\boldsymbol{h}_{T,k}$, (and similarly for the RIS-to-Rx channel) consisting of $M$ clusters. Each cluster contributes with one ray from the transmitter to the RIS. The ray parameters are: azimuth/elevation angles of arrival, $\theta_m, \phi_m \in [0,2\pi)$; complex coefficient $g_m \in \mathbb{C}$; time delay $\tau_m \in \mathbb{R}, \forall\, m \in \{1,2,\dots,M\}$. The transmitter-RIS path loss is denoted by $L_T$. The pulse shaping function, with $T_S$-spaced signaling, is defined as $p(\tau)$ at $\tau$ seconds. The delay-$d$ channel vector $\boldsymbol{h}_{T,d}$, can then be defined as

$$\boldsymbol{h}_{T,d} = \sqrt{\frac{N}{L_T}} \sum_{m=1}^{M} g_m p(dT_s - \tau_m) \mathbf{a}(\theta_m, \phi_m), \quad (4)$$

  where $\mathbf{a}(\theta_m, \phi_m) \in \mathbb{C}^{N\times 1}$ denotes the array response vector of the RIS at the angles of arrival $(\theta_m, \phi_l)$. Given this delay-$d$ channel, the frequency domain channel vector at subcarrier $k$, $\boldsymbol{h}_{T,k}$ can be expressed as

$$\boldsymbol{h}_{T,k} = \sum_{d=0}^{D-1} \boldsymbol{h}_{T,d} e^{-j\frac{2\pi k}{K} d}. \quad (5)$$

  We consider a block-fading channel model, $\boldsymbol{h}_{T,k}$ and $\boldsymbol{h}_{R,k}$ are assumed to be constant over the channel coherence time, denoted $T_c$, which depends on the mobility of the users and the dynamics of the environment. Hence, the reflection coefficient matrix $\boldsymbol{\Theta}$ only needs to be updated after every coherence interval $T_c$. It is worth noting that for mmWave channels, the value of $M$ can be very low whereas for sub-6 GHz signal propagation generally experiences rich scattering resulting in channels with more $M$.

## III. Problem Formulation

As stated previously, our main goal is to design the RIS reflection beamforming vector $\mathbf{v}$, to maximize the achievable rate at the receiver. Given the system and channel models in Section II, this achievable rate can be written as

$$R = \frac{1}{K}\sum_{k=1}^{K} \log_2\left(1 + \frac{P_T}{K\sigma_n^2}\left|\left(\boldsymbol{h}_{R,k} \odot \boldsymbol{h}_{T,k}\right)^T \boldsymbol{v}\right|^2\right). \quad (6)$$

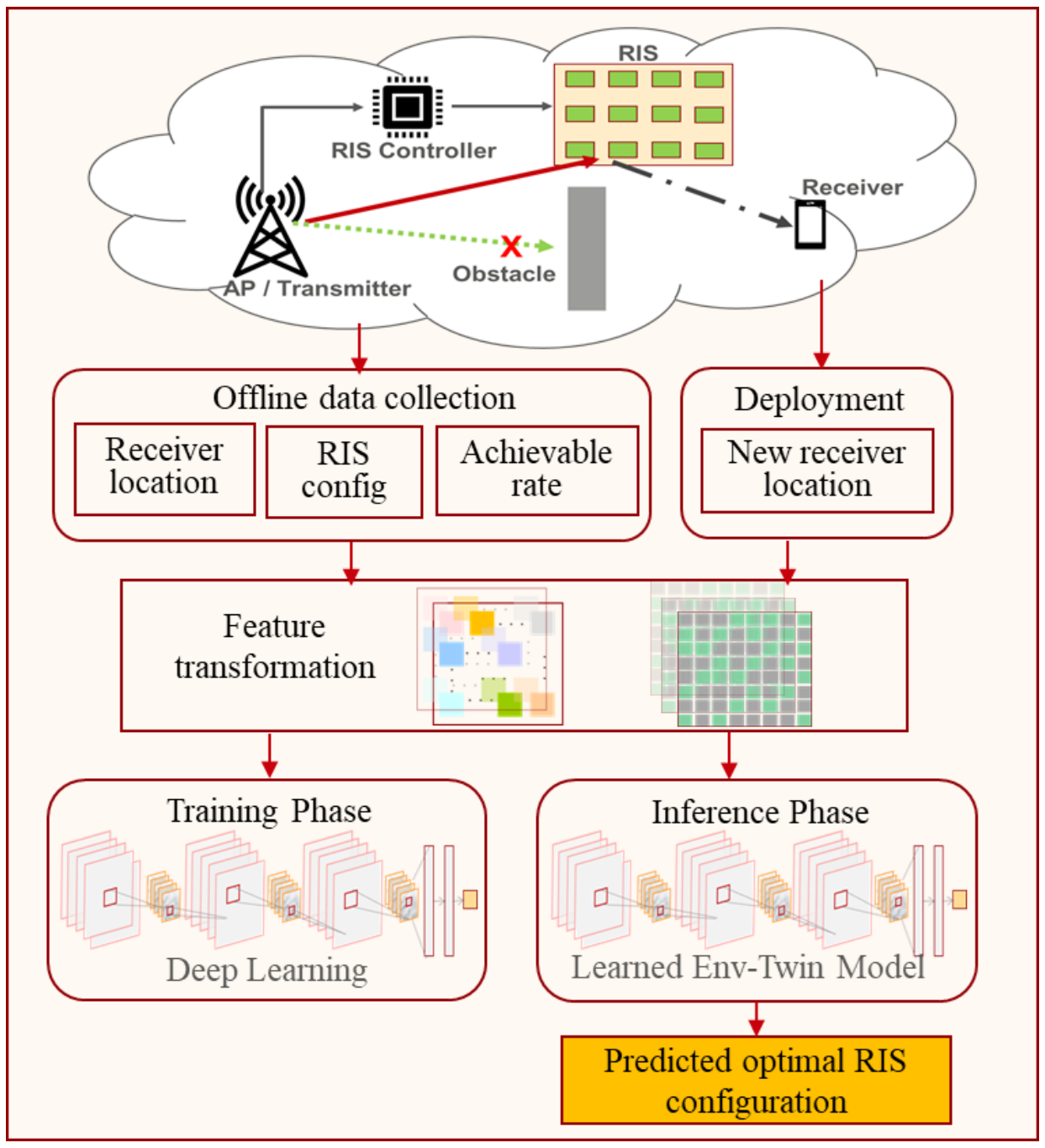

Fig. 2. Proposed solution framework for Env-Twin model learning and inference phases

In this paper, we also assume that the RIS elements can only take one of the discrete quantized set of angles due to hardware constraint [13][14]. Hence, we consider that the reflection beamforming vector $\mathbf{v}$ can only be picked from a pre-defined codebook $\boldsymbol{\mathcal{F}}$ . Each codeword in $\boldsymbol{\mathcal{F}}$ is assumed to be implemented using quantized phase shifting capability of RIS elements. Hence, our main goal is to find the optimal reflection beamforming vector $\boldsymbol{v}^*$ that satisfies the following

$$\boldsymbol{v}^* = \underset{\boldsymbol{v} \in \boldsymbol{\mathcal{F}}}{\operatorname{argmax}} \sum_{k=1}^{K} \log_2\left(1 + \frac{P_T}{K\sigma_n^2}\left|\left(\boldsymbol{h}_{R,k} \odot \boldsymbol{h}_{T,k}\right)^T \boldsymbol{v}\right|^2\right). \quad (7)$$

The above solution will provide the optimal rate $R^*$, which can be expressed as

$$R^* = \frac{1}{K}\sum_{k=1}^{K} \log_2\left(1 + \frac{P_T}{K\sigma_n^2}\left|\left(\boldsymbol{h}_{R,k} \odot \boldsymbol{h}_{T,k}\right)^T \boldsymbol{v}^*\right|^2\right). \quad (8)$$

Due to the quantized codebook constraint, there is no closed form solution for the optimization problem in (8). Moreover, such problems are non-convex optimization problems [12][14][15], and since the discrete phase shifts are constrained in a finite set $\boldsymbol{\mathcal{F}}$, the optimal solution can be obtained by the exhaustive search (ES) over the codebook $\boldsymbol{\mathcal{F}}$. It is worth noting that the size of the codebook should normally be in the same order of $N$, which means that for RIS with large $N$, ES will not be a feasible approach.

A traditional analytical solution would follow similar framework as above, which faces many challenges as discussed in Section I. Instead, in this paper we aim to find the optimal RIS reflection beamforming vector efficiently without any explicit channel estimation. We use (6)-(8) for the construction of observed achievable rates and the ground truth of optimal configurations for different receiver locations. Unlike [1][12], we assume no presence of any active RIS element. In the next section, we introduce our novel DL-based method to learn an Env-Twin model that maps the environmental attributes for a given location with any intended RIS reflection beamforming vector to the corresponding achievable rate by exploiting the location attributes of the possible receiver locations in the network.

## IV. Deep Learning-based Env-Twin Model

In this section, we introduce one example model that belongs to the Env-Twin framework as introduced in Section I. The purpose of this Env-Twin model is to learn the mapping function between the Rx location with any RIS reflection beamforming vector and the corresponding achievable rate without requiring explicit channel estimation effort. We leverage DL-based technique to build our Env-Twin model. First, we explain the key ideas and intuition behind the proposed method. Then, we describe our method in detail, and discuss how we construct the neural network architecture in the proposed method.

### A. *Key Ideas*

RIS-assisted wireless networks consist of Tx (e.g. wireless base station), Rx (e.g. user equipment), and RIS panel. From [16], the captured power at the Rx can be represented as some function of gains of RIS elements, wavelength, and Tx/Rx locations. However, with all the elements on the RIS being passive, antenna gain information at the RIS elements is not directly available whereas we understand it depends on RIS phase-shift configuration. With the intuition derived from [16], we design our ML-based approach to directly predict the achievable rate at a given Rx location after applying the RIS reflection beamforming vector $\mathbf{v}$ as described in (6) by using its location attributes and the intended $\mathbf{v}$, while bypassing explicit channel estimation. Our approach utilizes a DL-based approach to learn the mapping function given its strength in learning complex non-linear relationships. Our proposed method for modeling the RIS-embedded environment belongs to one of the models in the overall Env-Twin framework as described in Section I. For simplicity, we will use the term "Env-Twin" approach/model throughout this article. Once the Env-Twin-based model learns the mapping function, the optimal RIS configuration for any new Rx location can be obtained from the model's prediction results. This means the RIS-assisted communication network can approach optimal achievable rate as described in (8) without channel estimation overhead. Once trained, the Env-Twin model can be deployed as a software component at the controller which can be collocated with the RIS.

### B. *Proposed Method*

The proposed solution framework involves two phases in operation, namely the Env-Twin model offline training phase using training locations and the model inference phase for new Rx locations as depicted in Fig. 2.

1) *Model Training Phase:* During this phase, the system collects location measurement data offline from a set of sampled Rx locations, a set of RIS phase-shift configurations used in the data transmission for the sampled locations, and the corresponding achievable rates at the Rx locations. Suppose measurements for $U$ Rx locations are collected, each can be denoted as $\boldsymbol{l}_i = (l_{i,x}, l_{i,y}, \boldsymbol{d}_i)$ with $i = 1, 2, ..., U$, whereas $l_{i,x}$, $l_{i,y}$ represent Rx $i$'s $x$ coordinate, $y$ coordinate, and $\boldsymbol{d}_i$ represents Rx $i$'s distance to each of the elements on the RIS panel. Note that the size of $\boldsymbol{d}_i$ is equal to number of elements on the RIS, i.e. $N$. Part of the input for each training sample is RIS reflection beamforming vector for Rx location $i$. Each Rx location has $N$ possible RIS reflection beamforming vectors denoted as $\boldsymbol{v}_{i,p}$ with $p = 1, 2, \ldots, N$, where $N$ is the total number of RIS reflection beamforming vectors specified in the predefine codebook $\boldsymbol{\mathcal{F}}$. The default size of $N$ is $N_1 \times N_2$, where $(N_1, N_2)$ is the dimension of RIS, and for convenience of discussion we can set $N_1 = N_2$. Thus, the training input set for location $i$ can be denoted as

$$S_i \triangleq \{(\boldsymbol{l}_i, \boldsymbol{v}_{i,1}), (\boldsymbol{l}_i, \boldsymbol{v}_{i,2}), \ldots, (\boldsymbol{l}_i, \boldsymbol{v}_{i,N})\}.$$

The corresponding labeled output, represented by $R_{i,p}$ (bps/Hz) is the observed achievable rate at Rx location $i$ after RIS reflects using reflection beamforming vector $\boldsymbol{v}_{i,p}$, thus, there are a total of $N$ prediction achievable rate prediction results for Rx location $i$, one for each input $\boldsymbol{v}_{i,p}$ as defined in the codebook $\boldsymbol{\mathcal{F}}$.

Once the samples are acquired, the DL model is trained using these collected training samples. The training process allows the model to learn how to map an input sample (RIS reflection beamforming vector, $\boldsymbol{v}_{i,p}$, and location attributes $\boldsymbol{l}_i$ for Rx $i$) to its corresponding output (achievable rate, $R_{i,p}$).

2) *Model Inference Phase:* During the model inference phase, the system first receives estimated location attributes from the intended Rx $j$, namely its $x$ coordinates, $y$ coordinates, and calculates its distance features $\boldsymbol{d}_j$ to each of the RIS elements to form input $\boldsymbol{l}_j = (l_{j,x}, l_{j,y}, \boldsymbol{d}_j)$. Then, the system constructs input samples for location $j$ using all possible RIS reflection beamforming vectors specified in the predefined codebook $\boldsymbol{\mathcal{F}}$. The new input sample set for the intended Rx location $j$ can be denoted as

$$S_j \triangleq \{(\boldsymbol{l}_j, \boldsymbol{v}_{j,1}), (\boldsymbol{l}_j, \boldsymbol{v}_{j,2}), \ldots, (\boldsymbol{l}_j, \boldsymbol{v}_{j,N})\}.$$

After the new sample set is constructed, the trained DL model takes the input and predicts the achievable rates at location j for each possible RIS reflection beamforming vectors, $\boldsymbol{v}_{j,p}$, which can be expressed as

$$\hat{\boldsymbol{r}}_j \triangleq \{\hat{R}_{j,1}, \hat{R}_{j,2}, \ldots, \hat{R}_{j,N}\}.$$

The system performs an exhausted search across the achievable rates from the DL prediction output to determine the optimal RIS reflection beamforming vector for location $j$. This predicted optimal reflection beamforming vector $\hat{\boldsymbol{v}}_j$ is then used in data transmission for location $j$.

---

**Algorithm 1**: Deep Learning based Env-Twin Model

**Step 1: Model Training**

// Training Input: $Q$ (number of samples), receiver location features $\boldsymbol{l}_i \triangleq (x_i, y_i, \boldsymbol{d}_i)$,
// $i = 1, \ldots, U$, RIS reflection beamforming vector $\boldsymbol{v}_{1..p}$, $N_1$, $N_2$ (RIS dimension)
// Training Output: achievable rate $\hat{R}$ at receiver location $i$ after RIS reflects using $\boldsymbol{v}_p$

Input_train ← reshape(Input, ($Q$, $N_1$, $N_2$, dim(location attributes) + 2))
Output_train ← reshape(Output, ($Q$, 1))
$D$ ← (Input_train, Output_train)
Build CNN model $M$ for learning
Initialize weights, θ for $M$
**for** ep = 0 **to** epoch-1 **do**:
  Minibatch training dataset $D$
  forward propagation to calculate $\hat{R}$
  gradient descent on MSE($R$, $\hat{R}$) to obtain new θ
  update model weights using new θ
**end**

**Step 2**: **Model Inference**

// Input: trained CNN model $M$, location features $\boldsymbol{l}_j$ for intended receiver location $j$,
// predefined RIS reflection beamforming vectors $\boldsymbol{v}_{1..p}$
// Output: predicted best reflection beamforming vector for intended receiver location $j$

Input ← ( ($\boldsymbol{l}_j$, $\boldsymbol{v}_1$), ($\boldsymbol{l}_j$, $\boldsymbol{v}_2$), …, ($\boldsymbol{l}_j$, $\boldsymbol{v}_p$) )
Input ← reshape(Input, ($p$, $N_1$, $N_2$, dim(location attributes)+2))
$\hat{\boldsymbol{r}}$ = $M$.*predict*(Input)
*opt_index* ← argmax($\hat{\boldsymbol{r}}$)

RIS uses $\boldsymbol{v}_{opt_index}$ as the reflection beamforming vector for the intended receiver location $j$

---

The proposed DL-based operation is described in Algorithm 1.

## C. *Deep Neural Network Architecture and Parameters*

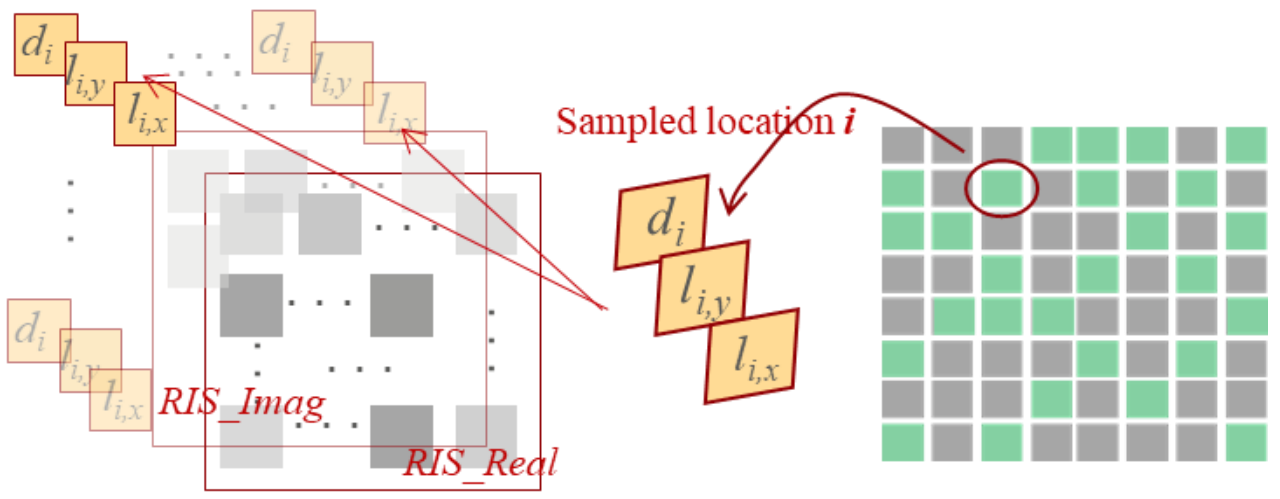

Fig. 3. This figure shows the representation of input features to the proposed Env-Twin CNN model

1) *Feature Representation:* Each input sample to the DL model comprises of RIS configuration, and Rx's location attributes. Inspired by [16, eq. (4)], we construct input features for each sample to the DL model, $(\boldsymbol{l}_i, \boldsymbol{v}_{i,p})$, as separate 2D channels. The first channel is the real part of $\boldsymbol{v}_{i,p}$, the second channel is the imaginary part of $\boldsymbol{v}_{i,p}$, and channels 3 – 5 are the location attributes. This representation is depicted in Fig. 3.

2) *Neural Network Architecture:* Our proposed DNN (deep neural network) architecture of the Env-Twin for RIS-assisted wireless networks is illustrated in Fig. 4. The input layer is a 5-channel map as described in the Feature Representation subsection, and the dimension of each channel is $N_1 \times N_2$, where we assume $N_1 = N_2$. The first channel comprises of the real part of the input RIS reflection beamforming vector, the second channel comprises of the imaginary part of the RIS reflection beamforming vector, the third channel comprises of the Rx location's x-coordinate, the fourth channel is the Rx location's y-coordinate, and the fifth and last channel comprises of the distance between the Rx location and each RIS element. Layers 2, 4, and 6 are convolutional layers, and each is followed by a pooling layer. The last pooling layer is followed by two fully connected layers (FCLs), and there is a dropout layer after each FCL. The last layer is a regression layer which predicts the scalar achievable rate for the input Rx location. We train the Env-Twin model using Adam optimizer [17] with default beta_1 and beta_2 and use 0.001 for the learning rate. During training, we set epochs to 100 and applied early stopping with patience set to 20, which ceases training procedure when the validation loss does not improve in 20 consecutive epochs.

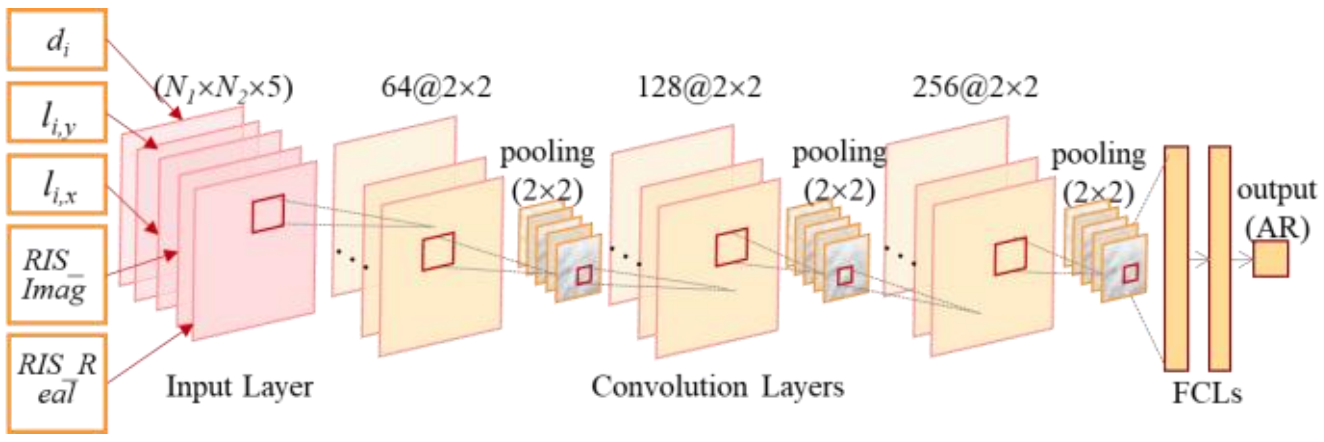

Fig. 4. The CNN Architecture for the proposed Env-Twin Model.

3) *Loss Function:* The objective of the Env-Twin is to accurately learn the mapping function and the output is the achievable rate for a given Rx location $i$ after applying RIS reflection beamforming vector $\boldsymbol{v}_{i,p}$. We train the Env-Twin model using a regression loss function of mean-squared-error (MSE), denoted as MSE ($\hat{R}$ , $R$) for each minibatch, in each epoch in the model training phase to minimize the MSE between the achievable rate predicted by the convolution neural network (CNN) model and the ground-truth label of achievable rate, which is obtained through simulation.

## V. Numerical Results

### A. Experimental Setup

In this section, we evaluate our performance by leveraging the public dataset, DeepMIMO [18]. This dataset includes scripts and tools that enables generating different dataset to simulate different scenarios by using specified parameters and settings, which facilitates ML research and development for mmWave/massive MIMO. Some research works [1][19] have already been published using this open dataset, thanks to its open-source and ease of use. To verify our approach, we also leverage the DeepMIMO ray-tracing scenario to generate sample data for training and testing our proposed Env-Twin model. Considering the use case of RIS and the goal of using RIS to reflect the beamforming, we adopt one base station as the RIS surface. To best simulate the impact of the environmental geometry on the realistic channels, ray-tracing is adopted to capture the dependence on the key environmental factors such as the environment geometry and materials the RIS and Tx/Rx locations, the operating frequency, etc. More specifically in this project, we chose the scenario O1_28 as the simulated real physical environment. The layout of scenario O1_28 is shown in Fig. 5.

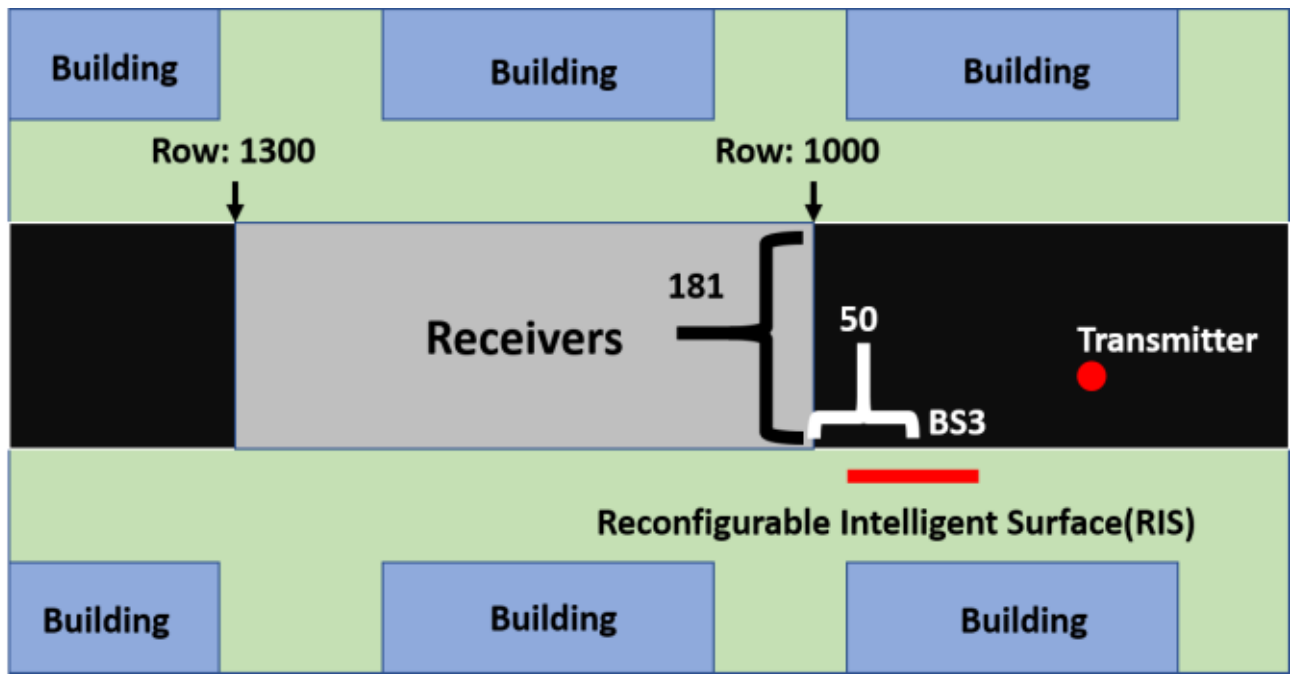

Fig. 5. This figure illustrates the DeepMIMO simulation scenario 'O1' used to generate the dataset for performance study of the proposed approach. The Tx is fixed at R950 and column 90. Candidate receiver is located between R1000 and R1300.

In the scenario O1_28, there is one Tx as labeled by a red dot in Fig. 5. The Rx's location area is shown by a gray rectangle starting from Row 1,000 to Row 1,300. For each row in the receiving area, there are 181 Rx locations. In another way, there are a total of $300 \times 181 = 54{,}300$ possible Rx locations in total in this scenario. Between the Tx and Rx's, BS3 is selected as the RIS for reflecting the beamforming, which is denoted by a red line.

TABLE I
The Adopted DeepMIMO Dataset Configuration Parameters

| Parameter | Value |
|---|---|
| Number of RIS Antennas | (*Nx;Ny;Nz*) = (1; 16; 16) |
| Antenna spacing | 0.5 |
| System bandwidth | 100 MHz |
| Number of OFDM subcarriers | 512 |
| OFDM sampling factor | 1 |
| OFDM limit | 16 |
| Number of paths | 5 |
| Transmit Power $P_T$ | 5 dB |

The configuration of the RIS and the communication parameters in O1_28 are shown in Table I. The RIS employs a uniform planar array (UPA) with 16x16 (256) antennas at the 28GHz setup. The Tx and Rx are assumed to have a single antenna each. Based on the above configuration parameters, the DeepMIMO simulation tools generate the RIS reflection beamforming codebook. In this scenario, it adopts a discrete Fourier transform (DFT) codebook [20][21] for the candidate RIS reflection beamforming vectors, which are also part of the input to the neural network of our method. Next, the DeepMIMO dataset used the ray-tracing simulator, Remcom Wireless InSite [22], to calculate the achievable bit rate for each location of the Rx based on the obtained DFT codebook and parameters for scenario O1_28. Given the above RIS configurations and parameters, there are a total of 256 reflection beamforming vectors in the codebook by default. The ray-tracing simulator calculates the achievable bit rate for each RIS reflection beamforming vectors for each location, which is treated as the ground truth of our neural network training process. Note that the DeepMIMO dataset doesn't contain separate location coordinates for each element on the RIS / BS3, thus, we use BS3's coordinates as feature input in our model.

## B. Evaluation Metrics

This subsection discusses the metrics we use for evaluating the performance of our work. Previous research [19] using the DeepMIMO dataset has demonstrated promising results using ML-based approach compared with compressed sensing-based approach. In [19], sampled channel vectors are used as input to predict achievable rates for all the RIS reflection beamforming vectors specified in the codebook at the same time. We, however, take a different approach. As described in Section I, one of our main goals is to reduce the beam training overhead by not relying on channel estimation. Instead, our approach only uses each Rx's location attributes with any RIS reflection beamforming vector configured in the pre-defined codebook $\boldsymbol{\mathcal{F}}$ as input to predict the corresponding achievable rate. To facilitate comparison with previous research, we use the same parameters to run the baseline model as described in [19] and our proposed approach. Note that requires active elements on the RIS panel and we set that to 8 ($\bar{M} = 8$) in our experiment to generate the baseline result. For performance evaluation, we use the following metrics:

- Top 1 prediction accuracy: This metric calculates the percentage of testing locations with correct prediction for the optimal RIS reflection beamforming vector, *i.e.* the predicted matches the ground truth according to the DeepMIMO ray-tracing scenario. The top 1 prediction accuracy, denoted as $Top1_{acc}$ can be expressed as

$$Top1_{acc} = \frac{1}{U}\sum_{i=1}^{U}\left(\widehat{Opt}(i) == Top_1{}^*(i)\right), \quad (9)$$

where $U$ is total number of testing Rx locations, $Top_1{}^*(i)$ is the true optimal RIS reflection beamforming vector for location $i$ according to the DeepMIMO ray-tracing simulation scenario, and $\widehat{Opt}(i)$ is the best RIS reflection beamforming vector for location $i$ according to the prediction results from the model.

- Top 3 prediction accuracy: This metric measures the percentage of testing locations whose predicted optimal RIS reflection beamforming vectors from the Env-Twin model are among the top 3 RIS reflection beamforming vectors according to the DeepMIMO ray-tracing simulation scenario. Note that RIS reflection beamforming vectors for each testing location are ranked in descending order according to the resulted achievable rates based on the DeepMIMO ray-tracing output which assumes perfect channel knowledge. Top 3 prediction accuracy, denoted as $Top3_{acc}$ can be expressed as

$$Top3_{acc} = \frac{1}{U}\sum_{i=1}^{U}(\widehat{Opt}(i) \in \{Top_1{}^*(i), Top_2{}^*(i), Top_3{}^*(i)\}), \quad (10)$$

where $U$ is total number of testing Rx locations, $Top_1{}^*(i)$ is the true optimal RIS reflection beamforming vector for location $i$, and $Top_2{}^*(i)$ and $Top_3{}^*(i)$ are the second and third best RIS reflection beamforming vector predictions for location $i$. $Top_1{}^*(i)$, $Top_2{}^*(i)$, and $Top_3{}^*(i)$ are obtained from the DeepMIMO ray-tracing simulation scenario.

- Recovered achievable rate percentage: For each testing location, we calculate the achievable rate reached when RIS applied the best reflection beamforming vector as predicted by the Env-Twin model. This predicted achievable rate is divided by the optimal achievable rate for the location based on the DeepMIMO dataset. We then average the results across all the testing locations to get the average recovered achievable rate percentage. The average recovered achievable rate percentage, denoted as *Recov_AR_avg*, can be expressed as

$$Recov_AR_avg = \frac{1}{U}\sum_{i=1}^{U}\frac{\widehat{AR}(i)}{AR^*(i)}, \quad (11)$$

where $U$ is total number of testing Rx locations, $\widehat{AR}(i)$ is the achievable rate after applying the best RIS reflection beamforming vector for location $i$ based on the prediction result of the Env-Twin model, and $AR^*(i)$ is the achievable rate for location $i$ after applying the true optimal RIS reflection beamforming vector based on the DeepMIMO ray-tracing simulation scenario.

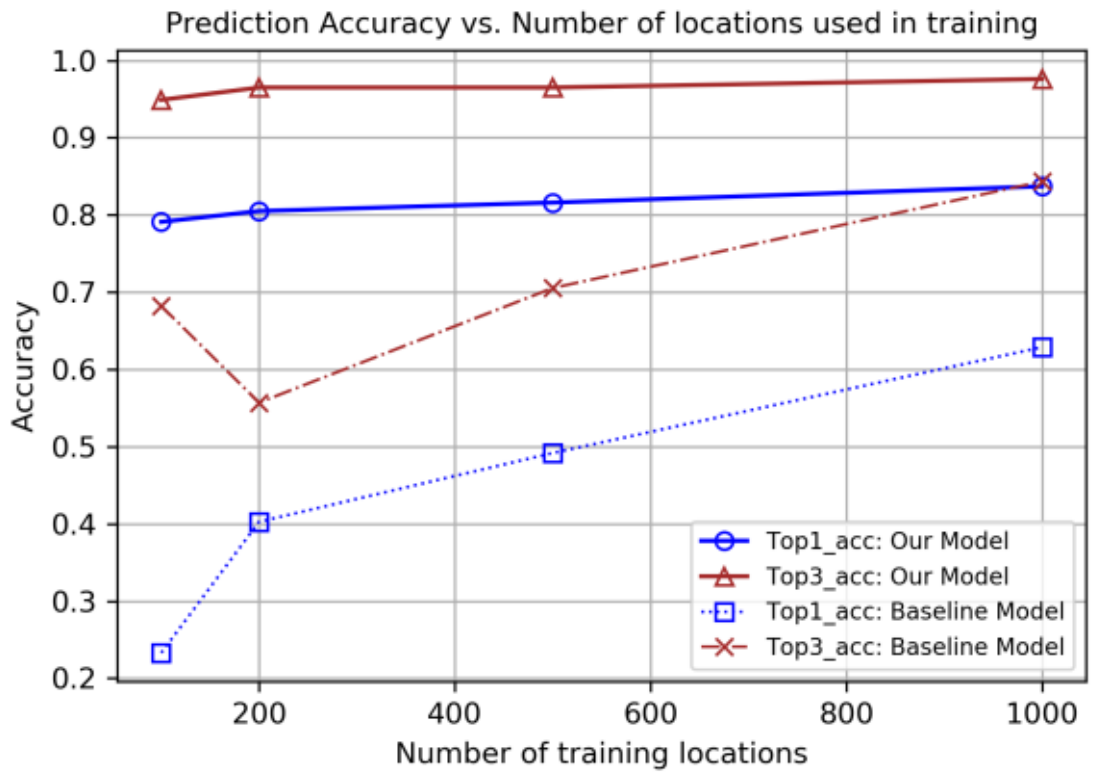

Fig. 6. This chart illustrates top 1 and top 3 prediction accuracy comparison between our approach and the baseline approach proposed in [19]. Results from each model are averages from 3 runs.

## C. Simulation Results

In this subsection, we discuss the performance evaluation results. We trained our model based on the generated DeepMIMO dataset for scenario O1_28 and evaluated the results using the metrics of average achievable rate and the Top $P$ accuracy, where $P \in \{1, 3\}$ as defined in the Evaluation Metrics subsection.

TABLE II
RECOVERED ACHIEVABLE RATE FOR TESTING LOCATIONS

| NUMBER OF TRAINING LOCATIONS | AVERAGE ACHIEVABLE RATE (PREDICTED) | RECOV_AR_AVG | AVERAGE OPTIMAL ACHIEVABLE RATE |
|---|---|---|---|
| 100 | 0.266 | 0.80 | 0.330 |
| 200 | 0.272 | 0.82 | |
| 500 | 0.288 | 0.87 | |
| 1000 | 0.289 | 0.88 | |

Fig. 6 illustrates the optimal RIS reflection beamforming vector prediction performance of the proposed Env-Twin model in terms of $Top1_{acc}$ and $Top3_{acc}$ as defined in (9) and (10), respectively. As shown, prediction accuracy improves when the number of training locations increases, and our proposed Env-Twin model achieved decent performance when using samples from only 100 Rx locations. We compared the performance with the approach proposed in [19], which also achieved decent top 3 accuracy when using 1,000 training locations, whereas the proposed Env-Twin model showed significantly better performance in top 1 accuracy when fewer training locations were used in training the model. Note that the result of each approach is the average of three independent runs of the corresponding method. We noticed performance degradation in the baseline method when number of training locations is increased from 100 to 200. This could be due to randomly initialized neural network weights.

To understand the achievable rate recovered after applying the best RIS reflection beamforming vector predicted by the model, we studied model performance by using different numbers of Rx locations in the training phase. We calculated *Recov_AR_avg* as defined in (11) and the results are described in Table II. We also compared the recovered achievable rate between our Env-Twin model, the approach proposed in [19] and random RIS configuration, and the results are illustrated in Fig. 7. As shown in Table II and Fig. 7, *Recov_AR_avg* increases when the number of training locations increases. Our model can achieve greater than 85% of the optimal rate when using only 500 locations (less than 1% of the total 54,300 locations) in training, which is significantly higher than both the baseline approach and random RIS configurations. With 1,000 training locations, our model can recover ~88% of the optimal achievable rate.

To further evaluate whether we can train a generalizable Env-Twin model using samples from a small percentage of the total Rx locations, we used our Env-Twin model that was built using 1,000 Rx locations (less than 2% of 54,300 locations) to predict the achievable rates for various numbers of testing receiver locations as illustrated in Fig. 8. Note that due to memory constraint on the machine, we ran predictions up to 21,000 testing receiver locations. As shown, the recovered achievable rate percentage stays stable at around 88% across all the testing location ranges. The above results suggest that our DNN-based Env-Twin solution is very sampling efficient, which enables practical realization of RIS control as it requires fewer number of training location samples.

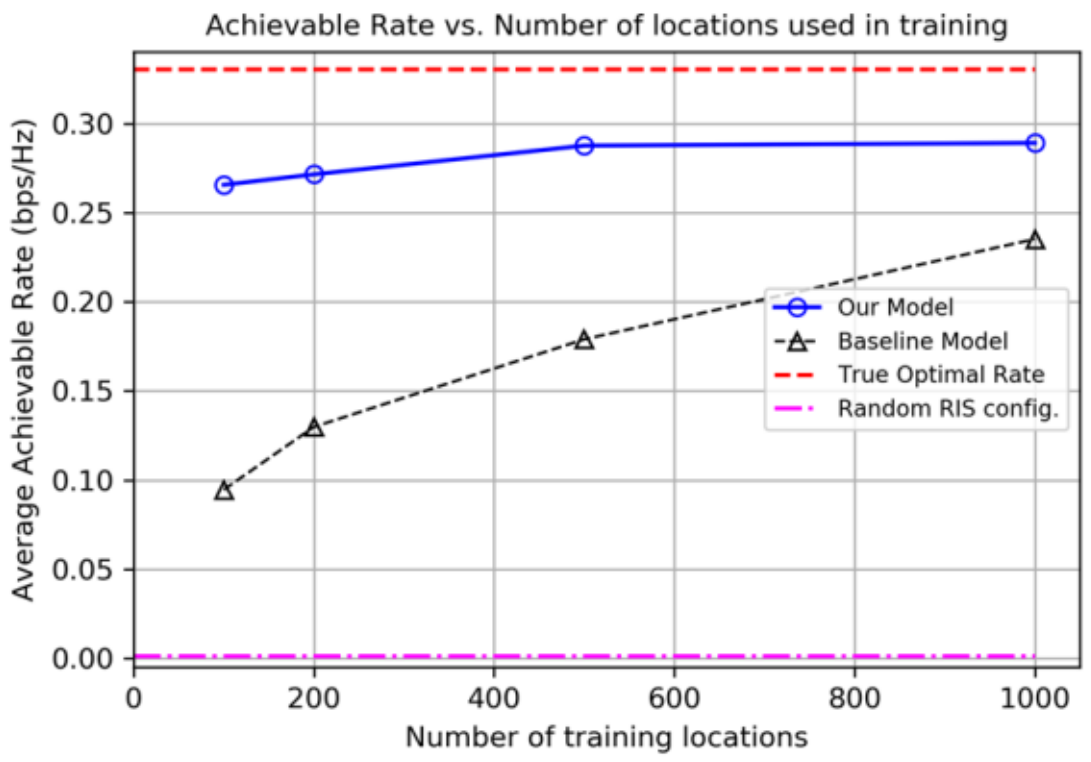

Fig. 7. The average achievable rate of our approach is compared with the baseline approach proposed in [19] and random RIS configuration. The upper bound is generated from the DeepMIMO dataset, which assumes perfect channel knowledge. Results for each approach are the averages from 3 runs.

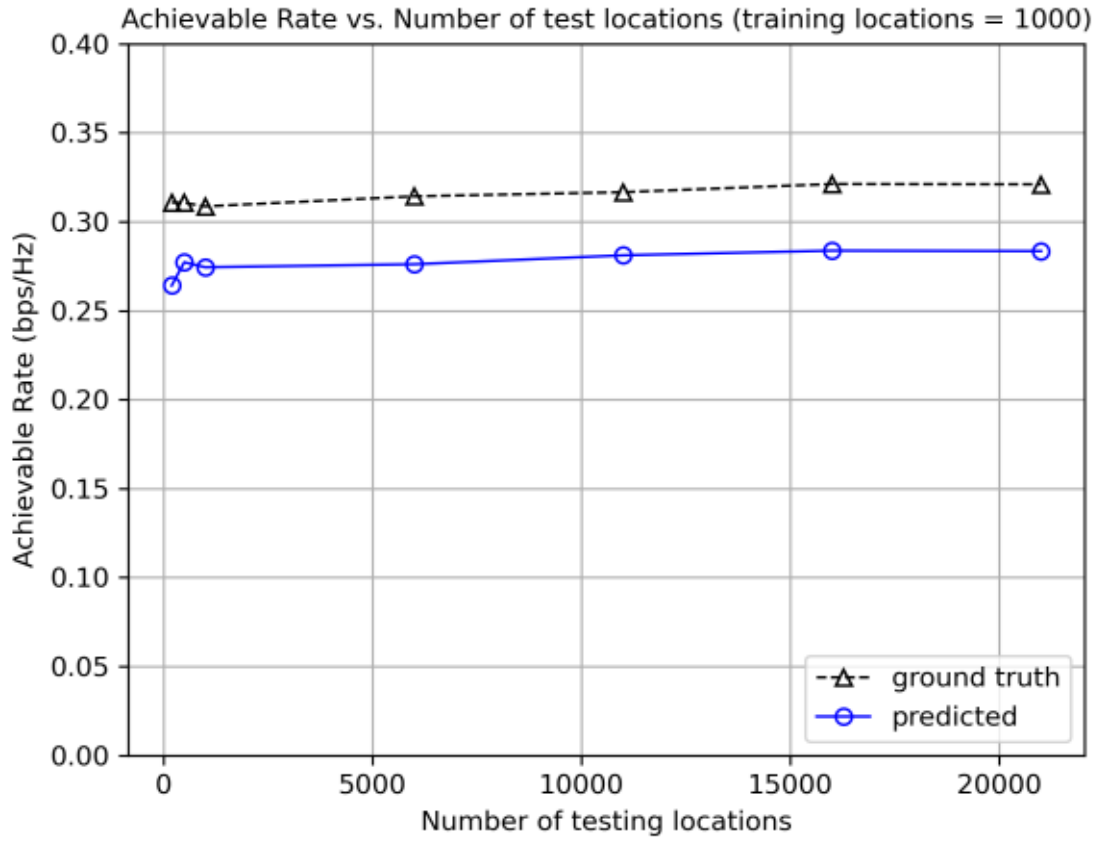

Fig. 8. Achievable rate using the predicted optimal RIS configuration vs. the true optimal achievable rate. Result shows prediction performance is stable when number of testing locations increases.

## VI. CONCLUSION

In this paper, we first introduced an Env-Twin framework for RIS-assisted wireless networks, which aims to capture the interactions and characteristics between the surrounding environment and the communication network and thus enable automation at various levels and granularities. Then, we developed one example of the Env-Twin models to learn the optimal RIS reflection beamforming vector directly from the location attributes of a given receiver without requiring any channel estimation effort. Our method leveraged communication domain knowledge as well as state-of-art DL techniques. Simulation results showed that our Env-Twin model can converge to near-optimal data rates using as little as 1,000 Rx locations during training, which is about 2% of the total number of possible receiver locations. This result demonstrates that our proposed solution is very sampling efficient, which means less overhead compared to approach using channel information estimated at the active elements of the RIS, as shown in the prior art. For future research, we plan to study the practical application challenges of the proposed method in a real (not simulation) environment. Meanwhile, how to combine ML-based approach with traditional analytical method is a potential research topic to leverage the strengths of both. Another interesting topic is to include other non-RF sensing inputs in the model learning stage, e.g. images from depth-camera to capture environmental objects in the communication networks.